\input{aipcheck}

\documentclass[
    ,final            
  ]
  {aipproc}

\layoutstyle{6x9}

\begin{document}
\include{def}
\def\brn{\begin{eqnarray*}}
\def\ern{\end{eqnarray*}}
\def\mbphi{\mbox{\boldmath$\Phi$}}
\def\a {{\alpha}}
\def\b {{\beta}}
\def\g {{\gamma}}
\def\qb {{\bf q}}
\def\piN {{\pi N}}
\def\mbtbig{\mbox{\boldmath$T$}}
\def\mbwbig{\mbox{\boldmath$W$}}
\def\mbomega{\mbox{\boldmath$\omega$}}
\def\mbrho{\mbox{\boldmath$\rho$}}
\def\mbzeta{\mbox{\boldmath$\zeta$}}
\def\Lh{\hat{\cal L}}
\def\qslash{/\!\!\!{\rm q}}
\def\pslash{/\!\!\!{{\rm p}}}
\def\kslash{/\!\!\!{{\rm k}}}
\def\Qslash{/\!\!\!{{Q}}}
\def\ppslash{/\!\!\!{{\rm p'}}}
\def\psidxbnu{\bar{\psi}_{\Delta \nu}(x)}
\def\psix{\psi_N(x)}
\def\n {\nu}
\def\s{\sigma}
\def\l {\lambda}
\def\g {\gamma}
\def\d {\dagger}
\def\e {\epsilon}
\def\Rb {{\bf R}}
\def\Tb{ {\bf T}}
\def\sss{\scriptscriptstyle}
\def\ss{\scriptstyle}
\def\bb {{\bf b}}
\def\E {{{\cal E}}}
\def\ninj#1#2#3#4#5#6#7#8#9{\left\{\negthinspace\begin{array}{ccc}
#1&#2&#3\\#4&#5&#6\\#7&#8&#9\end{array}\right\}}
\def\sixj#1#2#3#4#5#6{\left\{\negthinspace\begin{array}{ccc}
#1&#2&#3\\#4&#5&#6\end{array}\right\}}
\def\M {{{\cal M}}}
\def\sqi{\frac{1}{\sqrt{2}}}
\def\x{\times}
\def\nn{\nonumber }
\def\w {{\omega}}
\def\endauthors{}
\def\authors#1\endauthors{#1}
\def\fot{\frac{1}{2}}
\def\tfot{\frac{3}{2}}
\def\mbs{\mbox{\boldmath$\sigma$}}
\def\mbp{\mbox{\boldmath$\phi$}}
\def\mbpi{\mbox{\boldmath$\pi$}}
\def\mbt{\mbox{\boldmath$\tau$}}
\def\bin#1#2{\left(\negthinspace\begin{array}{c}#1\\#2\end{array}\right)}
\def\binv#1#2{\left\{\negthinspace\begin{array}{l}#1\\#2\end{array}\right.}
\def\rf#1{{(\ref{#1})}}
\def\ov#1#2{\langle #1 | #2  \rangle }
\def\sss{\scriptscriptstyle}
\def\ss{\scriptstyle}
\def\bra#1{\langle #1|}
\def\ket#1{|#1 \rangle}
\def\Ket#1{||#1 \rangle}
\def\Bra#1{\langle #1||}
\def\rb {{\bf r}}
\def\pb {{\bf p}}
\def\Pb{ {\bf P}}
\def\be{\begin{equation}}
\def\ee{\end{equation}}
\def\br{\begin{eqnarray}}
\def\er{\end{eqnarray}}
\def\binv#1#2{\negthinspace\begin{array}{c}#1\\#2\end{array}}
\def\cuav#1#2#3#4{\negthinspace\begin{array}{c}#1\\#2\\#3\\#4\end{array}}
\def\binn#1#2{\left\{\negthinspace\begin{array}{l}#1\\#2\end{array}\right.}
\def\jp{{\sf{j}}_p}
\def\jN{{\sf{j}}_N}
\def\jL{{\sf{j}}_\Lambda}
\def\js{{\sf{j}}}
\def\trinn#1#2#3{\left\{\negthinspace\begin{array}{l}#1\\#2\\#3\end{array}\right.}

\title{Single photon production $\nu_l N\rightarrow \nu_l N \gamma$ in neutrino-nucleon scattering}


\author{C. Barbero}{
  address={Departamento de F\'{\i}sica, Universidad Nacional de La Plata, C.
C. 67, 1900 La Plata, Argentina}
}

\author{A. Mariano}
{
  address={Departamento de F\'{\i}sica, Universidad Nacional de La Plata, C.
C. 67, 1900 La Plata, Argentina}
}


\begin{abstract}
The quasielastic charged current (CCQE) $\nu_e n \rightarrow e^- p$ scattering is the dominant mechanism to detect appearance
of a $\nu_e$ in an almost $\nu_\mu$ flux at the $1$ GeV scale.
Actual experiments show a precision below $1 \%$ and between less known background contributions, but necessary to constraint the event excess,
we have the radiative corrections.
A consistent model recently developed for the simultaneous description of elastic and radiative $\pi N$ scattering,
pion-photoproduction and single pion production processes, both for charged and neutral current neutrino-nucleon scattering, is
extended for the evaluation of the radiative $\nu_l N\rightarrow \nu_l N \gamma$ cross section.
Our results are similar to a previous (but inconsistent) theoretical evaluation in the low energy region, and show an increment in the upper region where the $\Delta$ resonance becomes relevant.
\end{abstract}

\maketitle


\section{Introduction}

Neutrino interactions with nucleons and nuclei have received a
considerable attention in recent years stimulated by the
needs in the analysis of neutrino oscillation experiments which give
information about the oscillation probability $P(\nu_i\rightarrow\nu_j)$.
Actually, new high quality data on cross sections are becoming available from MiniBooNE
and SciBooNE experiments \cite{Mini}. The neutrino energy that enters critically in the oscillation probability is not directly
measurable but has to be reconstructed from the reaction products. In appearance experiments the initial and final neutrinos fluxes
are compared to get the oscillation probability. In fact, MiniBooNE was designed to measure the
$\nu_\mu\rightarrow\nu_e$ appearance signal
in the energy range around $1$ GeV, detecting electrons from the $\nu_e n\rightarrow e^- p$ reaction via Cerenkov radiation. In this case,
neutral current $\nu_l N\rightarrow \nu_l N\pi^0$ events (NC$1\pi^0$) are an important source of background
because the electron is not distinguished from a photon coming from $\pi^0\rightarrow \gamma\gamma$
decay if the other photon is missed. This would lead to fake $\nu_e$ signal in contrast with the true coming form the CCQE.
However, another important source of single-photon background exists:
the $\nu_l N\rightarrow \nu_l N \gamma$ (NC$1\gamma$) reaction, which can also be interpreted as a $\nu_e$ arriving signal.

Recent MiniBooNE experiment reported, after various refinements, a persisting  excess of electron-like events  at low energy \cite{Arevalo09}. In the recent NC$1\gamma$ cross section  calculation and analysis from Hill \cite{Hill09}, it has been shown that process seems capable to provide enough photons to cover the excess found by MiniBooNE. However, the implemented model  is not completely {\it consistent} since the vertex and propagator used for the $\Delta$ correspond to a different value of the parameter associated to  contact transformations on the spin 3/2 field (see \cite{Bar08} and references therein). Then, it is important to treat consistently and on the same footing both, the $1\pi$ and photon production cross sections, to provide a trustily constraint of these of backgrounds.
In the $0-2$ GeV region relevant for MiniBooNE and SciBooNE experiments, the process following in importance to the QE scattering is the excitation of the $\Delta(1232$ MeV) resonance.
Nuclei are used as neutrino detectors and require the inclusion of nuclear
medium effects on the free nucleon cross sections.
A consistent treatment should remedy the before mentioned resonance vertex-propagator correspondence and also keep the preservation of the electromagnetic gauge invariance in presence of finite widths. This program has been already developed with success for another reactions involving the $\Delta$ resonance, which encourage us to develop here a model for the evaluation of  the cross section of the $\nu_l N\rightarrow \nu_l N \gamma$ reaction.

\section{Formalism}

\begin{figure}
  \includegraphics[height=.1\textheight]{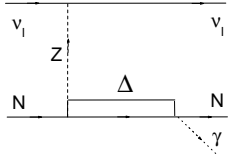}
  \vspace{-2.0cm}
  \caption{Resonant contribution to single photon production.}
\label{fig1}
\end{figure}

We perform here the calculation of NC$1\gamma$ cross section by considering only the contribution of the resonant amplitude shown in Fig. \ref{fig1}.
The total cross section is given by

\be
\sigma(E_\nu^{\sss
CM})= \frac{m_N^2}{(2\pi)^4E_\nu^{\sss CM}\sqrt{s}}
\int_0^{{E'_\nu}^{+}}
d{E'_\nu}^{\sss CM}
\int_{E_\gamma^{-}}^{E_\gamma^{+}} dE_\gamma^{\sss CM}
\int_{-1}^{+1}dcos\theta
\int_0^{2\pi}d\eta {1 \over 32}
\sum_{spin,pol}\left|{ \cal M}\right|^2,
\label{crossW}\ee
where $E_\nu^{\sss CM}=\frac{m_NE_\nu}{\sqrt{2E_\nu m_N+m_N^2}}$, with the limits in the integrals estimated from kinematical conditions, and where $E_\nu\equiv E^{LAB}_\nu$. The resonant amplitude reads

\[
{\cal M} \simeq {\cal M}_R=i\frac{G_F}{\sqrt{2}}\bar{u}({\rm p}_\nu)\g_\l(1-
\g_5){u}({\rm p}_\nu)\bar{u}({\rm p}'){\cal O}_i^\lambda({\rm p},{\rm p}',{\rm q})u({\rm p}),
\]
where ${\rm p}_\nu$, ${\rm p}$ and ${\rm p}'$ are the neutrino, initial and final nucleon momenta, respectively,
and ${\rm q}={\rm p}-{\rm p}_\Delta$ is the $Z$ boson momentum. ${\cal O}_R$ reads ($q_\gamma$ is photon momentum and $Q^2=-q^2$)
\br
{\cal O}_R^\lambda({\rm p},{\rm p}',{\rm q})= {\overline{\Gamma}}^{\alpha\nu}({\rm p}',{\rm p}_\Delta,q_\gamma={\rm p}_\Delta-{\rm p}')\epsilon_\nu^*
~iG_{\alpha\beta}({\rm p}_\Delta={\rm p - q}){\cal W}^{\beta\lambda}(p_\Delta,{\rm q}={\rm p}-{\rm p}_\Delta,{\rm p}),
\label{deltapole}\er
Some calculations involving the $\Delta$ production take the simplest
form for ${\overline{\Gamma}}^{\alpha\nu}$ (contact constant $A=-1$) \cite{Bar08}
and, simultaneously, the simplest one for the vertex ($A=-1/3)$. To avoid this problem and get $A$ -independent amplitudes we use a set of reduced $A$-independent Feynman rules \cite{Bar08}.
The expression for the corresponding bare $\Delta$ propagator and weak vertex ${\cal W}_{\beta\lambda} = {\cal W}^V_{\beta\lambda}-{\cal W}^A_{\beta\lambda}$ was defined in Ref. \cite{Bar08}.
Finally, the
$\Delta \rightarrow \gamma N $ decay vertex, from which ${\cal W}^V$ is extracted through the CVC hypothesis, was defined in Ref. \cite{Mariano07}. The amplitude \rf{deltapole}, should be multiplied  by
the new isospin factor
${\cal T}_R(m_tm_{t'})=\chi^\dagger(m_{t'})T_3(\mbtbig^\dagger\cdot {\bf Z^*})\chi(m_t)
=2/3$ for both processes with single photon production induced by neutral currents: $\nu_\mu p \rightarrow  \nu_\mu p \gamma$ and $\nu_\mu n \rightarrow  \nu_\mu n \gamma$.

\section{Numerical results and summarizing conclusions}

\begin{figure}
\vspace{-3cm}
  \includegraphics[height=.2\textheight]{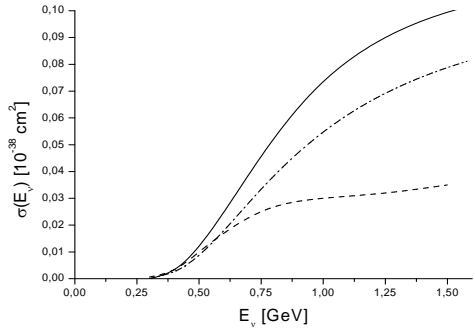}
  \vspace{-5.0cm}
  \caption{Cross section for NC$1\pi^0$ from Ref. \cite{Bar08} (solid line), NC$1\gamma$ ($\x 100$) background to CCQE scattering from Eq. (1) (dash-dot line) and from Ref. \cite{Hill09} (dash line).}
\label{fig2}
\end{figure}

We show in Figure \ref{fig2} the cross section for NC$1\pi^0$ and NC$1\gamma$ backgrounds to the CCQE process,
for $0\leq E_\nu\leq 1.5$ GeV. The contribution of the radiative
$\nu_l N\rightarrow \nu_l N\gamma$ process is roughly one hundred times smaller than the neutral current one, since  $g_{\Delta N\gamma}=\sqrt{4\pi/137}<<<f_{\Delta N\pi}=\sqrt{4\pi\x 0.317}$. In the same figure we compare our results with those from Fig. 4 in Ref. \cite{Hill09}, obtained within an inconsistent model. We can observe that differences are not important  at low energies but yes in the upper region where the $\Delta$ resonance becomes relevant. This is also an expectable result originated by the fact we are considering the correct complete $\Delta$ propagator ($A=-1/3$) in Eq. \rf{deltapole}.

In summary,
the results obtained within our present {\it consistent} formalism for
 $\nu_l N\rightarrow \nu_l N\gamma$ process were compared with a previous determination that lacks in consistence in the treatment of the resonance field \cite{Hill09}.
Our treatment gives similar results in the low energy region and increases the single photon cross section in the $\Delta$ region.
In view of these results, it would be interesting to reanalyze the excess of events counting contribution of this reaction done in Ref. \cite{Hill09}. This finally answer the question if radiative corrections provides enough photons to cover the low energy excess found by MiniBooNE.



\bibliographystyle{aipproc}   


\IfFileExists{\jobname.bbl}{}
 {\typeout{}
  \typeout{******************************************}
  \typeout{** Please run "bibtex \jobname" to optain}
  \typeout{** the bibliography and then re-run LaTeX}
  \typeout{** twice to fix the references!}
  \typeout{******************************************}
  \typeout{}
 }


\end{document}